\documentstyle[12pt]{article}
\begin{document}
%%%%%%%%%%%%%%%%%%%%%%%%%%%%%%%%%%%%%%%%%%%%%%%%%
\title{ 
\begin{flushright}
\small
MPI-PHE/96-02 \\ DFTT 1/96 \\[6ex]
\normalsize
\end{flushright}
Light Neutralinos as Dark Matter in the Unconstrained Minimal
Supersymmetric Standard Model\thanks{Submitted to Astroparticle Physics, 
19 Feb. 96}}

\author{
A. Gabutti$^{a}$\thanks{Corresponding author,
gabutti@vms.mppmu.mpg.de}, M. Olechowski$^{b,c}$,
S. Cooper$^{a}$, \\
S. Pokorski$^{a,c}$, L. Stodolsky$^{a}$ 
\\[0.2cm]
{\it \small $^{a}$ Max-Planck-Institut f\"ur Physik, F\"ohringer Ring 6, 
D-80805 M\"unchen, Germany}\\
{\it \small $^{b}$ INFN Sezione di Torino and Dipartamento di Fisica Teorica,}\\
{\it \small Universit\`a di Torino, Via P. Giuria 1, 10125 Turin, Italy}\\
{\it \small $^{c}$ Institute of Theoretical Physics, 
Warsaw University, ul. Hoza 69, 00-681 Warsaw, Poland} 
}

\date{ }
\maketitle
\begin{abstract}
The allowed parameter space for the lightest neutralino as the dark matter is
explored using the Minimal Supersymmetric Standard Model as the 
low-energy effective theory without further theoretical constraints 
such as GUT.
Selecting values of the parameters which are in agreement with present 
experimental 
limits and applying the additional requirement that the lightest neutralino 
be in a cosmologically interesting range, we give limits on the neutralino 
mass and composition. A similar analysis is also performed implementing
the grand unification constraints.
The elastic scattering  cross section of the selected 
neutralinos on $^{27}$Al and on other materials for dark matter experiments
is discussed.
\end{abstract}

\newpage

\section{Introduction}
$\quad$
Particle candidates for dark matter are classified as hot or cold depending 
on whether they were relativistic or not at the time they decoupled from 
thermal equilibrium.
Any particle candidate for cold dark matter implies an extension of the 
Standard Model.
At present the most interesting candidate is the lightest neutralino 
of the Minimal Supersymmetric Standard Model (MSSM).

Any weakly interacting massive particle (WIMP) considered as a dark matter
candidate is subject to at least two constraints:
1) its relic abundance must be cosmologically interesting, 
say $0.025 <  \Omega h^2 < 1$, in units of the critical density and
2) its existence must be in accord with 
present experimental limits, provided mainly by the LEP experiments.

In the standard approach, based in solving the kinetic Boltzman equation,
the relic abundance of WIMPs is given roughly by \cite{ref:Omega}:
	\begin{equation}\label{eq:relic}
\Omega h^{2} \approx \frac{10^{-37} \mbox{cm}^2 }
                          {\langle\sigma_{ann} v \rangle}
	\end{equation}
where $\langle\sigma_{ann} v \rangle$ is the thermal average at freezeout
of the annihilation cross section times the relative velocity in units of $c$.
This is a very remarkable result: $\Omega h^{2}\sim 1$ 
for typical weak cross sections.
Suppose the annihilation of a WIMP of mass $m$ proceeds
via the exchange of a particle of mass $M$ 
(where $M^{2}=x M_{Z}^{2}$)
coupled with the strength $g$ ($g^{2}=y g_{Z\nu \nu}^{2}$),
where we have introduced the scaling factors $x$ and $y$ 
to express the mass $M$ and the coupling $g$ in terms of 
the $Z$ boson mass and its coupling to neutrinos.
Using Eq.~(\ref{eq:relic}) and the expressions for 
$\langle\sigma_{ann} v \rangle$ which follow from dimensional arguments
we have:
\begin{equation}
\langle\sigma_{ann} v \rangle\ \sim\ 
    \frac{y^2}{x^2}\ \frac{m^2}{M_Z^4}\ \frac{g_{Z\nu\nu}^4} {16\pi^2}\ 
 =\ \frac{y^2}{x^2}\ \left(\frac{m}{1\,\mbox{GeV}}\right)^2
    \times 0.4 \times 10^{-37}\,\mbox{cm}^2\ 
    \qquad \mbox{for } m\ll M 
\end{equation}
$$ 
\langle\sigma_{ann} v \rangle\ \sim\ 
    y^2\ \frac{1}{m^2}\ \frac{g_{Z\nu\nu}^4} {16\pi^2}\ 
 =\ y^2\ \left(\frac{m}{1\,\mbox{TeV}}\right)^{-2}
    \times 0.2 \times 10^{-37}\,\mbox{cm}^2\ 
    \qquad \mbox{for } m\gg M
$$
and requiring $\Omega h^2 < 1$ leads to a generalized Lee-Weinberg bound of: 
	\begin{equation}\label{eq:lee}
\frac{x}{y} \times {\cal O}(\mbox{1\,GeV}) < m < 
y \times {\cal O}(\mbox{1\,TeV})\ .
	\end{equation}

This is a useful qualitative constraint on the mass of a WIMP 
when used together with experimental limits 
on the parameters $x$ and $y$ 
(of course $x$ and $y$ are also constrained 
by the requirement that $\Omega h^{2}$ should not be too small).
It is convenient to distinguish two cases: 
a neutrino-like WIMP which interacts only via $Z$-boson exchange and
a non neutrino-like WIMP which can annihilate 
via exchange of new particles.

In the first case, the presently available experimental constraint
on the ``invisible" width of the $Z$,
$\Delta\Gamma_{inv} / \Gamma_{\nu} < 0.05$~\cite{ref:Ginv}, 
gives $y < 0.05$ (for of course $x$=1 and $m < M_Z/2$). 
Therefore, from Eq.~(\ref{eq:lee}) we obtain 
$m > {\cal O}(\mbox{20\,GeV})$.
This is an order of magnitude estimate; in particular
we neglect the difference between Majorana and Dirac particles.

The second case is realized in the Minimal Supersymmetric Standard Model
with the lightest neutralino as the dark matter candidate. Here the
analysis is much more involved because of the complexity of the model
and its large parameter space.
A considerable amount of work has been devoted to the neutralino
as a dark matter candidate~\cite{ref:Griest}. 
However in most cases several additional theoretical assumptions are used
(such as radiative electroweak symmetry breaking 
and universal boundary conditions at the GUT scale for the 
parameters of the soft breaking of supersymmetry)
which may be too restrictive and certainly go beyond the MSSM as the 
low-energy effective theory.

In view of the dark matter search experiments,
we explore in Sec.~\ref{sec:neutra} the most general scenario 
for neutralinos as dark matter within the 
MSSM as the low-energy effective theory without any further theoretical 
constraints such as the grand unification constraints (GUT).
For completeness, 
the obtained neutralino masses and compositions are compared with the results
derived using the GUT constraints.

In Sec.~\ref{sec:det} we consider the direct detection of neutralino dark
matter via elastic scattering on nuclei. The dependence of the axial 
coupling (spin dependent) on the nuclear model and on the quark spin
contents is discussed for selected neutralino compositions.
In Sec.~\ref{sec:cross} we calculate the neutralino cross
section on some of the nuclei presently used or planned to be used for dark 
matter detection.
We analyze the low mass region and give the dependence of the  
cross section on the neutralino mass and composition.
Prospects for the detection of light neutralinos are discussed in 
Sec.~\ref{sec:CRESST}.

Our analysis differs from previous work on neutralino
detection \cite{ref:Griest} because our more general assumptions allow 
the possibility
of having low mass neutralinos as dark matter candidates and we evaluate their
cross section on nuclei.
In the present work
neutralinos with relic abundance below the cosmological 
bound $\Omega h^2 < 0.025$ are not considered as dark matter candidates.
This is different from the approach used in other work \cite{ref:Bottino} 
where neutralinos with relic abundance below the cosmological bound are still
considered as dark matter candidates and their contribution to the
dark matter halo is evaluated by rescaling the local dark matter density.

\section{Neutralino in the MSSM as dark matter candidate}
\label{sec:neutra}
$\quad$
In the present section we address the question of the most general limits on 
the lightest neutralino mass and its composition 
which follow only from the two constraints $0.025 < \Omega h^2 < 1$ 
and consistency with the present data from accelerator experiments,
without any further theoretical assumptions.
As we shall see, with this approach it is possible 
to obtain quite strong qualitative conclusions.

The stable neutralino is the lowest mass superposition of
neutral gauginos and higgsinos:
	\begin{equation}\label{eq:neut}
\chi_1 = Z_{11} \tilde{B}     + Z_{12} \tilde{W}^{0} +
         Z_{13} \tilde{H}_1^0 + Z_{14} \tilde{H}_2^0
	\end{equation}
or, in the photino, zino, higgsino basis:
        \begin{equation}\label{eq:neut2}
\chi_1= a \tilde{\gamma}    + b \tilde{Z} +
        Z_{13}\tilde{H}_1^0 + Z_{14}\tilde{H}_2^0
	\end{equation}
with 
$\tilde{\gamma} = \cos \theta_W \tilde{B} + \sin \theta_W \tilde{W}^0$, 
$\tilde{Z}    = - \sin \theta_W \tilde{B} + \cos \theta_W \tilde{W}^0$ and
$\theta_{W}$ the Weinberg angle.
The neutralino composition is defined by the neutralino mass matrix,
which in the basis of Eq.~(\ref{eq:neut}) is:
\begin{equation}
\left( \begin{array}{cccc}
M_{1} &0 &-M_{Z}\cos\beta \sin\theta_{W} &M_{Z}\sin\beta \sin\theta_{W} \\
0 &M_{2} &M_{Z}\cos\beta \cos\theta_{W} &-M_{Z}\sin\beta \cos\theta_{W} \\
-M_{Z}\cos\beta \sin\theta_{W} &M_{Z}\cos\beta \cos\theta_{W} &0 &-\mu \\
M_{Z}\sin\beta \sin\theta_{W} &-M_{Z}\sin\beta \cos\theta_{W} &-\mu &0 \\
\end{array} \right)
\end{equation}

\subsection{Scan of the parameter space}
\label{subsec:scan}
$\quad$
The neutralino mass matrix depends on several, in general free, parameters of 
the model: the gaugino masses $M_1$ and $M_2$, the higgsino mass parameter 
$\mu$, and $\tan\beta$=$v_2/v_1$, where $v_1$ and $v_2$ are 
the vacuum expectation values of the two Higgses present in the model. 
All of these are independent parameters of the low-energy effective lagrangian.
In the present study of the unconstrained low-energy effective theory, 
the values of these parameters are chosen randomly
in the ranges listed in Tab.~\ref{tab:range}.
The scanning procedure is such that $\log(M_{2})$, $\log(\mu)$, 
$\log(\tan\beta)$
and the ratio $M_{1}/M_{2}$ are chosen with flat probability.
We restrict our analysis to neutralinos lighter than 100\,GeV. 

Calculations with the grand unification constraints are also performed using 
the same ranges but with $M_1$=$\frac{5}{3}$M$_{2}\tan^{2}\theta_{W}\simeq$
0.5$M_2$ and with $M_{2}\simeq$0.3$m_{\tilde{g}}$. 
We use for the gluino mass $m_{\tilde{g}}$ the experimental lower bounds 
from CDF \cite{ref:CDF} and D0 \cite{ref:D0} resulting in $M_{2}\ge$50 GeV.
It should be noted that the present experimental limits on the gluino
mass are derived only for a specific choice of the MSSM parameters.

We scan the parameter space incorporating the following 
existing experimental limits from accelerator experiments. 
1) The limit on the $Z$ boson invisible width is
$\Gamma_{inv} < 8.4$\,MeV~\cite{ref:Ginv}. 2)
Heavier neutralinos are unstable and could be observed in $e^{+}e^{-}$ 
collisions via their decay products
for the range of $\chi$ masses which are kinematically accessible.
We use the mass-dependent limits on the coupling constants 
given in Ref.~\cite{ref:ALEPHRep}.
3) The limit on the chargino mass is 
$M_{\chi^{\pm}} > 45$\,GeV~\cite{ref:ALEPHRep}.
The chargino mass matrix is a function of the same parameters:
\begin{equation}
\left( \begin{array}{cc}
M_{2} & \sqrt{2} M_{W} \sin\beta \\
\sqrt{2} M_{W} \cos\beta & \mu\\
\end{array} \right) 
\end{equation}
and does not depend on the ratio $M_{1}/M_{2}$.

The scanned parameter space is shown in Fig.~\ref{fig:M2mu} with and
without grand unification constraints after incorporating the limits from
accelerator experiments.
The empty regions in the $M_{2}-\mu$ plane are excluded by the 
experimental limits.
The quoted chargino mass limit is at present by far the strongest 
constraint on the parameter space as shown in Fig.~\ref{fig:M2mu} where the
contour line $M_{\chi^{\pm}} = 45$\,GeV is plotted
for $\tan \beta$=1 and $\tan \beta$=50.
Without grand unification constraints, the region corresponding to small 
values of $M_{2}$ and $|\mu|$ is excluded for $\tan \beta > 3$.

\subsection{Relic abundance}
\label{subsec:relic}
$\quad$
We then calculate the relic neutralino abundance in the allowed parameter 
region. We solve the Boltzman equation following 
Ref.~\cite{ref:gelmini}, using a complete set of supersymmetric annihilation 
amplitudes~\cite{ref:program}. 
For this we need the squark, slepton, and Higgs masses.
We assume all squarks have the same mass $M_{sq}$ and all sleptons the same 
mass $M_{sl}$.
Since the purpose of our study is to explore the 
parameter space, we need limits which are
model-independent, whereas most experimental limits are
based on certain assumptions on the values of the parameters.
Thus we have chosen to repeat our calculation for several
different values of the squark, slepton and CP-odd Higgs masses as
 shown in Tab.~\ref{tab:set}.
In the case of sets A--C, the upper bound in the neutralino mass is 
given by the slepton mass since we require $M_{\chi}\le M_{sl}$.

The lowest values of the squark masses are taken as 45\,GeV, 
since lighter particles are ruled out by LEP data on the Z.
CDF~\cite{ref:CDF} has shown that the squark mass limits 
obtained from $p\bar p$ collisions are dependent on the assumed squark decay 
modes and that for many configurations of the MSSM parameters  
the squark mass is unconstrained.
A recent report by D0~\cite{ref:D0} gives limits on the squark mass
as function of the gluino mass for a particular set of MSSM parameters,
assuming that all squarks have the same mass.
Even for this particular set of parameters, squarks can be as light as 
$\sim$50\,GeV for $m_{\tilde{g}}\geq$300\,GeV.
OPAL~\cite{ref:OPAL} has set limits of about 45\,GeV specifically for the 
$\tilde t$ mass, which may be significantly lighter than the other squarks 
due to the left-right mixing. 

For sets A--C the CP-odd Higgs mass $M_A$ is randomly chosen in the range 
25-70\,GeV in the regions of the plane $M_{A}$-$M_{h}$ allowed by present 
LEP data \cite{ref:OPALhiggs}. 
In sets D and E, $M_{A}$ is fixed at 200\,GeV and 1 TeV respectively.
The other Higgs masses are calculated from $M_A$ and $\tan\beta$ using
radiative corrections \cite{ref:radcorr} which introduce a dependence on
the top quark mass $M_{t}$ and $M_{sq}$. The top quark mass is set to 170\,GeV.

The result of our calculation for the relic abundance $\Omega h^2$ 
obtained without the grand unification constraints are plotted in
Fig.~\ref{fig:omega} for sets B--E. The cosmologically relevant neutralinos
are selected imposing the condition $0.025 < \Omega h^2 < 1$.
It must be noted that we do not rescale the local dark matter density for 
neutralinos with relic abundance below the cosmological bound 
$\Omega h^2 < 0.025$ since we do not consider them as dark matter
candidates.

For low squark and slepton masses (set B) the condition $\Omega h^2 < 1$ 
is obtained for neutralino masses above 2\,GeV.
Decreasing the squark masses to 45\,GeV (set A) gives a relic abundance very
similar to the one of set B and the same value 
of 2\,GeV for the lowest neutralinos, which can then be regarded as
an absolute lower bound on the neutralino mass in the MSSM.
This is one of the main conclusions of this study.
The lower bound in the neutralino mass rises with $M_{sq}$ and
$M_{sl}$ to reach $\sim$20\,GeV in sets D and E.

The relic abundance evaluated using the grand unification constraints is plotted
in Fig.~\ref{fig:omegaGUT} for sets B--E.
Here neutralinos have a lower bound on the mass of 20\,GeV,
due to the experimental constraints on the gluino mass.
A relic abundance $\Omega h^2 << 1$ is obtained for sets A (not shown) and B 
where we use low squark and slepton masses.

\subsection{Neutralino composition}
\label{subsec:comp}
$\quad$
In Fig.~\ref{fig:purity}  we show the gaugino purity
$Z_{11}^2 + Z_{12}^2$ and separately the bino purity $Z_{11}^2$ for the 
cosmologically relevant neutralinos derived without grand unification 
constraints combining data sets A--E.
The points at bino purity $Z_{11}^2 = \cos^{2} \theta_{W} \simeq$ 0.77
 correspond to pure photinos.
It is seen that the cosmologically relevant neutralinos 
are dominantly gauginos (mostly bino and photino), 
with a substantial higgsino component coming in above 30\,GeV.
The neutralino composition calculated with GUT is shown in 
Fig.~\ref{fig:purityGUT} and it is similar to 
the composition obtained without grand unification constraints except that 
the photino population of the gaugino-like neutralinos is strongly reduced.
The higgsino composition is shown in Fig.~\ref{fig:Z3Z4} without grand 
unification constraints. 
Since we consider values of $\tan\beta>$1 we have 
$Z_{13}^{2} > Z_{14}^{2}$ with  $Z_{13}=-Z_{14}$ when $\tan\beta$=1.
The higgsino composition with grand unification constraints is similar to the
one shown in Fig.~\ref{fig:Z3Z4}. 

The depopulation in the region with large gaugino-higgsino mixing is due
to the scanning of the allowed parameter space and to the
requirement of having cosmologically relevant neutralinos.
The neutralino is a mixed state only when the smaller of the two gaugino 
masses, $M_{2}$ or $M_{1}$, has a magnitude similar to $|\mu|$. 
This is a fine-tuning which is not included in our scanning procedure, in 
consequence we have a substantial reduction in 
the population of mixed neutralinos.
Especially for large $M_{2}$ there are few points in the parameter space
corresponding to mixed neutralinos. In this case, for values of
$M_{1}$ not much smaller than $M_{2}$, mixing is allowed for
large values of $|\mu|$ giving neutralinos heavier than our upper
bound in the neutralino mass of 100\,GeV. The only mixing allowed
for large $M_{2}$ corresponds to both $M_{1}$ and $|\mu|$ being small.

The dependence of the relic abundance on the neutralino composition is shown 
in Fig.~\ref{fig:omegapur} for sets B and D  without grand 
unification constraints.
Pure gauginos do not have couplings to the $Z$ boson and have higher relic
abundance than pure higgsinos. The relic abundance of pure photinos
is bigger than that of pure bino because the former does not couple to
the Higgs bosons.
At large $\tan\beta$ the annihilation is dominated
by Higgs exchange which is proportional
to the gaugino-higgsino mixing and goes to zero for pure states.
As a result, large gaugino-higgsino mixtures  have low relic abundance and
are excluded by the request of having $\Omega h^2 > $ 0.025.
The dependence of the cosmologically relevant neutralino compositions 
on $\tan\beta$ is 
shown in Fig.~\ref{fig:omegatan} for sets B and D. 
Gaugino-higgsino mixtures are relevant only at small $\tan\beta$ while
at large $\tan\beta$ our cosmologically relevant neutralinos are mainly pure 
gauginos with exception of set E which has a very heavy CP-odd Higgs boson.
There is a substantial photino component for $\tan\beta <$2 while at large 
$\tan\beta$ the photino component is strongly reduced due to the structure
of the neutralino mass matrix.
Due to the large squark mass, photino have relic abundance $\Omega h^2>1$ 
in the case of sets D and E.

\section{Neutralino detection}
\label{sec:det}
$\quad$
Dark matter neutralinos can be directly detected via 
their elastic scattering on nuclei.
The scattering cross section has two components.
An effective axial-vector interaction gives a spin-dependent (SD) 
cross section which is non-zero only for nuclei with net spin.
Scalar and vector interactions give spin-independent (SI) cross sections
which involve the squares of nuclear neutron and proton numbers.
The relative strengths of the two parts 
of the neutralino interaction on nuclei 
depends on the neutralino composition.
A pure gaugino couples only to squarks and sleptons and has only a SD
interaction. The sleptons are not relevant for interactions with nuclei.
A pure higgsino couples only to the axial-vector part of the 
$Z$ boson and thus has only a SD interaction.
(The Yukawa coupling to squarks is negligible for scattering 
on the proton.)
Mixed states have both SD and SI interactions, the later is due to squark
exchange and Higgs exchange and is proportional to the zino-higgsino mixture.

The actual scattering cross section depends on the coefficients
$Z_{1j}$ that specify the neutralino composition, and 
on the masses $M_{sq}$ and $M_A$ of the exchanged particles.
To obtain the neutralino scattering on nuclei for each set of supersymmetric 
parameters, we first calculate the coefficients of the effective 4-fermion 
operators describing neutralino-quark scattering.
These operators are used to give the neutralino-nucleon amplitudes,
which are then related to the nuclear matrix elements.

We shall focus on calculating  the cross section for neutralino scattering on a 
nucleus at zero momentum transfer. This incorporates the essentials of
the physics of neutralino cross section on nuclei. In Sec.~\ref{sec:CRESST}
we will include in the evaluation of the recoil energy spectrum and 
detection rate nuclear form factor effects due to the momentum transfer. 
In the following, we will call the integral
of the zero momentum transfer cross section over all momentum transfer
 $\sigma_{0}$.
The cross section $\sigma_{0}$ is derived from: 
        \begin{equation}\label{eq:sigma0}
\sigma_{0}=
4 \frac{G_{F}^{2}}{\pi} \Big[S_{A}(0)+S_{S}(0)\Big] M_{red}^{2}
        \end{equation}
where  
the effective axial-vector current $S_{A}(0)$ and the scalar current 
$S_{S}(0)$ coefficients are calculated at zero momentum 
transfer ($q$=0) and $M_{red}$ is the reduced mass.
The coefficients $S_{A}$ and $S_{S}$ are evaluated using the
expressions derived in Ref.~\cite{ref:x3}.

\subsection{Spin Dependent cross section}
\label{subsec:SD}
$\quad$
The evaluation of the SD cross section is problematic because it depends 
on the model used to describe the spin structure function of the nucleon
and on the nuclear model used to derive the total proton and
neutron spins.
In this section we discuss the resulting uncertainty in the determination
of the axial-vector current coefficient:
\begin{equation}\label{eq:odmodel}
S_{A}(0)=8 J(J+1) \bigg\{\frac{a_{p}<S_{p}>}{J}+
\frac{a_{n}<S_{n}>}{J}\bigg\}^{2}
\end{equation}
with:
\begin{equation}
a_{p(n)}=
\Big[\sum_{u,d,s}^{}A_{q}\Delta Q_{q}\Big]_{p(n)} 
\end{equation}
where $J$ is the nuclear spin and $<S_{p(n)}>$ are the total proton and neutron
spins. The terms $A_{q}$  refer to the coupling
of the up, down and strange quarks weighted with the
nucleon quark spin content coefficients $\Delta Q_{q}$. 
The entire dependence on the nature of the neutralino, and therefore of
our cosmological considerations, enters only through the coefficients
$A_{q}$. For the determination of $<S_{p(n)}>$ and $\Delta Q_{q}$, we
rely entirely on previous work \cite{ref:NQM}-\cite{ref:shellge}.

It is illustrative to consider pure photino and pure bino interactions 
for which the proton axial coupling coefficients $a_{p}$ are \cite{ref:x3}:
\begin{eqnarray}\label{eq:ap}
&&a_{p}=-\frac{M_{W}^{2}}{M_{sq}^{2}} \tan^{2}\theta_{W}
\bigg\{\frac{17}{36}\Delta Q_{u}
+\frac{5}{36}(\Delta Q_{d}+\Delta Q_{s})\bigg\} \qquad \mbox{pure bino} \\
&&a_{p}=-2\frac{M_{W}^{2}}{M_{sq}^{2}}\sin^{2}\theta_{W}
\bigg\{
\frac{4}{9}\Delta Q_{u} + \frac{1}{9}(\Delta Q_{d}+\Delta Q_{s})\bigg\} 
\qquad \mbox{pure photino} \nonumber 
\end{eqnarray}
Since the quark spin coefficients $\Delta Q_{q}$ are evaluated for scattering on
a proton, the neutron axial coupling coefficient $a_{n}$ is obtained by 
interchanging the values of $\Delta Q_{u}$ and $\Delta Q_{d}$ according to the
different quark structure of the two nucleons.

In Tab.~\ref{tab:DQ} we have calculated $a_{p(n)}$ from Eq.~(\ref{eq:ap})
using for $\Delta Q_{q}$ the values derived
from the Naive Quark Model (NQM) \cite{ref:NQM}, the 
European Muon Collaboration (EMC) measurements \cite{ref:EMC} 
and from two analyses of the 
present data on polarized lepton-nucleon scattering
Global Fit-1 from Ref.~\cite{ref:nucspin} and the more recent 
Global Fit-2 from Ref.~\cite{ref:nucspin2}.
The axial coupling coefficients $a_{p}$ are roughly a factor of two
larger than $a_{n}$ due to the different quark structure of the two nucleons
which enhances the axial coupling on protons and therefore on
proton-odd nuclei.

The dependence of the axial-vector current coefficient $S_{A}(0)$ on the model 
used to describe the nuclear structure is shown in Tab.~\ref{tab:Sa} 
where we have calculated the axial-vector 
current coefficient for pure photino and pure bino interactions on the 
proton-odd nucleus $^{27}$Al and on the neutron-odd nucleus $^{73}$Ge.
We compare the predictions from the Odd Group Model (OGM) \cite{ref:OGM} 
and the detailed Shell Model reported in Ref.~\cite{ref:MICA} for 
$^{27}$Al and in Ref.~\cite{ref:shellge} for $^{73}$Ge.
We also show the effects of different model for the spin quark coefficients
$\Delta Q_{q}$.
  
If the OGM is used to describe the nuclear structure, only 
the unpaired nucleon contributes to the axial coupling. In this case the 
axial-vector current coefficients $S_{A}(0)$ are proportional
to $a_{p}^{2}$ for proton-odd nuclei and $a_{n}^{2}$ for neutron-odd nuclei. 
In the OGM, where only the valence nucleon plays a role, the cross sections on
nuclei obtained with the NQM are roughly a factor of two larger on proton-odd 
nuclei than those for other quark models directly reflecting the
cross sections on single nucleons.
On the contrary, the SD cross section on neutron-odd nuclei derived 
with the NQM is more than an order of magnitude smaller than the one
obtained using the other $\Delta Q_{q}$ values due to the small value of 
$a_{n}$ in the NQM.

It should be noted that the values for the
total proton and neutron spins derived in the detailed Shell Model 
calculations for $^{27}$Al have an uncertainty of roughly
30$\%$ related to the quenching of the spin matrix elements.
Quenching is more important in
$^{73}$Ge and we use the quenching factor $Q=0.833$ in the isovector piece of
the axial coupling coefficients $a_{p(n)}$ as suggested in 
Ref.~\cite{ref:shellge}.
In this case the axial coupling coefficients have to be replaced
with $a_{p}^{\prime}=0.917a_{p}+0.0835a_{n}$ and 
$a_{n}^{\prime}=0.0835a_{p}+0.917a_{n}$ where $a_{p(n)}$ are evaluated in the
unquenched case (Eq.~(\ref{eq:ap})).
The cross section on $^{27}$Al evaluated using the detailed Shell 
Model calculations are roughly 2 times 
larger than the results obtained with the OGM for all the four model
used to describe the spin structure of the nucleon.
The detailed Shell Model gives almost 4 times larger
SD cross sections on $^{73}$Ge than the OGM for all the three quark models 
with non zero strange quark content. 
The axial current coefficients for pure photinos and binos are smaller
for interactions on $^{73}$Ge than on $^{27}$Al mainly because of the different
quark structure of neutrons and protons.

In general neutralinos are mixtures of gauginos and higgsinos.
For pure higgsinos, the axial coupling coefficient $a_{p}$ derived
from Ref.~\cite{ref:x3} is:
\begin{eqnarray}\label{eq:higgs}
a_{p}&=&\frac{1}{4}\Big(Z_{13}^{2}-Z_{14}^{2}\Big) \Big[\Delta Q_{u}-
\Delta Q_{d}-\Delta Q_{s}\Big]\\
&&-\frac{1}{2} \frac{1}{\sin^{2}\beta} \frac{1}{M_{sq}^{2}}
\bigg\{Z_{14}^{2}M_{u}^{2}\Delta Q_{u} + Z_{13}^{2} \tan^{2}\beta 
\Big[M_{d}^{2}\Delta Q_{d} + M_{s}^{2}\Delta Q_{s}\Big]\bigg\} \nonumber
\end{eqnarray}
where $M_{u,d,s}$ are the quark masses. The neutron axial coupling coefficient
$a_{n}$ is obtained by interchanging the values of
$\Delta Q_{u}$ and $\Delta Q_{d}$.
For the same higgsino composition, the difference between $a_{p}$ and $a_{n}$
is mainly due to the first term of Eq.~(\ref{eq:higgs}) which is always positive
for protons and negative for neutrons since $Z_{13}^{2} \ge Z_{14}^{2}$ 
(see Fig.~\ref{fig:Z3Z4}).
The second term of Eq.~(\ref{eq:higgs}) does not change significantly going from
$a_{p}$ to $a_{n}$ because it is dominated by the contribution of the strange 
quark which is much heavier than the $u$ and $d$ quarks.
As a result, $|a_{p}| > |a_{n}|$  and higgsino interactions have the largest 
cross section on neutron-odd nuclei.

\subsection{Spin Independent cross section}
\label{subsec:SI}

$\quad$
We evaluate the scalar current coefficient $S_{S}$ from:
        \begin{equation} \label{eq:SS}
S_{S}(0)= 2 A^{2}
\big[P_{s}\lambda_{s}+\sum_{c,b,t}^{}P_{h}\lambda_{h}\big]^{2}
        \end{equation}
where $A$ is the atomic number. The terms $P_{i}$ are determined by the 
couplings in the effective $\chi \chi q \tilde{q}$ lagrangian, with
contributions from squark exchange and Higgs exchange of
the lightest and the heaviest Higgs bosons.
The supersymmetric content of the interaction is represented by the
$P_{i}$ terms.
The coefficients $\lambda_{i}$ are given by the matrix elements of the
quark operators $\frac{M_{i}}{M_{W}} q_{i} \tilde{q}_{i}$ taken 
between the nucleon states and are by standard methods \cite{ref:Griest}
expressed in terms of the sigma-term measured 
in pion-nucleon interactions and the nucleon mass.
For $\lambda_{i}$ and 
$P_{i}$\footnote{The original notation of 
Ref.~\cite{ref:x3} is $S_{q}$ instead of $P_{i}$}
we use the values given in Ref.~\cite{ref:x3}.

The contribution of the light $u$ and $d$ quarks to the scalar 
neutralino-nucleon coupling can be 
neglected and the term $P_{i}\lambda_{i}$ accounts for the coupling to the 
strange quark (index $s$) and to the heavy quarks $c$, $b$ and $t$ (index $h$).
Since we do not consider the contribution of the $u$ and $d$ quarks, 
we do not have to distinguish between scalar cross section on neutron
or proton.

It must be noted that the SI cross section evaluated using the 
$P_{i}\lambda_{i}$ terms of Ref.~\cite{ref:x3} 
is four times smaller than the SI cross section derived in 
Ref.~\cite{ref:Bottino}. This factor of four is independent of the neutralino
composition and mass.

\section{Cross section on nuclei}
\label{sec:cross}
$\quad$
In this section we derive the SD and SI cross sections on some of the
material presently used or planned to be used for dark matter detection.
In contrast to our illustrative examples in the previous section, we will now
consider arbitrary neutralino compositions, as given by our scan for
cosmologically relevant cases.

\subsection{$^{27}$Al}
\label{subsec:Al}
$\quad$
We start with the proton-odd nucleus $^{27}$Al (100$\%$ natural abundance).
Dark matter detectors made of sapphire ($Al_{2}O_{3}$) are presently under 
preparation by the CRESST \cite{ref:CRESST} and 
EDELWEISS collaborations \cite{ref:EDELWEISS}. 

In order to evaluate the SD cross section  
we chose to use the quark spin coefficients $\Delta Q_{q}$ derived
in Ref.~\cite{ref:nucspin2} (Global Fit-2) and 
the results of the detailed Shell Model calculations reported
in Ref.~\cite{ref:MICA} for the nuclear physics.
In Fig.~\ref{fig:SISD} the SI cross section is plotted versus the SD part 
for the cosmologically relevant neutralinos without the grand
unification constraints. 
There is a 5 orders of magnitude spread in the values of the SD cross section
for the neutralino masses and compositions considered in our analysis. 
The spread is even more pronounced in the case of the SI cross section.

The dependence of the SI cross section on the neutralino composition 
is shown in Fig.~\ref{fig:BinoSI} for data sets B and D and $\tan\beta <$2.
Since the scalar coupling is proportional to the zino-higgsino
mixture, pure neutralino states have zero SI cross section.
Mixtures with a large photino component have very small SI
cross section $< 10^{-6}$ which is almost 5 orders of magnitude
smaller than the corresponding SD part.
The cross section is reduced to values
below $10^{-3}$ pbarn for binos and higgsinos with purity $>$ 0.99.

The dependence of the SI cross section on $\tan\beta$ is shown in
Fig.~\ref{fig:SItanbeta}. 
Although the SI cross section is expected to increase with increasing 
$\tan\beta$, the maximum value of the SI cross section for our selected 
neutralinos changes only slightly with $\tan\beta$. 
This is related to the dependence of the neutralino composition on $\tan\beta$
as discussed in Sec.~\ref{subsec:comp}.
At small $\tan\beta$ gaugino-higgsino mixtures are cosmologically relevant 
giving the maximum SI cross section.
At large $\tan\beta$  our selected neutralinos are mainly bino 
or higgsino with a reduced SI cross section.

The SD and SI cross sections obtained without the grand unification 
constraints are plotted versus the neutralino mass in Fig.~\ref{fig:crossAl}
for the combined data sets A--E.
The bands of points in the low mass region of the SD cross section 
are due to pure photinos and pure binos as shown in 
Fig.~\ref{fig:photcross}.
The SD cross section of pure binos is $\sim$2 times smaller than the one
for pure photinos as shown in Tab.~\ref{tab:Sa}.
The photino contribution to the SD cross section is
relevant only for small $\tan\beta$ because the photino population 
in our selected neutralinos is strongly reduced at high $\tan\beta$.
The highest SD cross section is obtained for the pure gauginos of set A.
In this case squark exchange gives a significant 
contribution because of the small squark mass $M_{sq}$=45\,GeV.

In Fig.~\ref{fig:BinoSD} the SD cross section is plotted versus 
the bino purity for sets B and D without grand unification 
constraints. For heavy squarks (set D) the highest SD cross section is 
obtained for gaugino-higgsino mixtures with a substantial higgsino component. 
This is because squark exchange is suppressed and $Z$ exchange becomes dominant.
  
The SD cross section for our neutralinos is sometimes 
even larger than the cross section for a massive Majorana neutrino interacting
via Z exchange;
this happens when the squark mass is low and its exchange 
gives a significant contribution to the cross section.
In contrast, the maximum values of the SI cross section 
are two orders of magnitude below that for Dirac neutrinos.

The SI and SD cross sections evaluated with the grand unification
constraints are  plotted in Fig.~\ref{fig:crossGUT} for the cosmologically 
relevant neutralinos of the combined data sets A--E.
The main contribution to the cross section is given by sets C--E since for
sets A and B the neutralino relic abundance is $\Omega h^{2} <<$ 1 
(see Fig.~\ref{fig:omegaGUT}).
The horizontal bands in the SD cross section are due to the bino contributions
of sets C and D since in the case of GUT
the photino population is suppressed as shown in Fig.~\ref{fig:purityGUT}.
Comparing the cross sections obtained with and without grand unification
constraints, we see that the maximal value of $\sigma_{0}$ and the
dependence of the cross section on the neutralino
mass are almost the same. In the case of GUT, 
 the cutoff in the neutralino mass is at $\simeq$20\,GeV as discussed in 
Sec.~\ref{subsec:relic}.

\subsection{$^{23}$Na and $^{127}$I}
$\quad$
We turn now to the evaluation of the neutralino cross section on the 
proton-odd nuclei $^{23}$Na and $^{127}$I both with 100$\%$ natural abundance.
NaI(Tl) scintillator detectors are presently used for dark matter searches 
 by several groups \cite{ref:dmNaI}. 

The cross sections $\sigma_{0}$ (SI+SD) on $^{23}$Na and on $^{127}$I are 
plotted in Fig.~\ref{fig:crossNa} 
for the combined sets A--E without grand unification constraints. 
Since $^{27}$Al and $^{23}$Na have similar atomic numbers 
the SI cross sections in the two materials do not differ significantly.
Due to the difference in the atomic numbers, the SI cross section on $^{127}$I
is roughly two orders of magnitude larger than on $^{23}$Na.

The dependence of the SD cross section on the neutralino composition
is the same for Na, I and Al because these nuclei have an unpaired
proton. 
The dependence of the SD cross section on the absorber material is given
by the reduced mass, the nuclear spin and the total proton and neutron
spins $<S_{p(n)}>$. 
In the case of $^{23}$Na and $^{127}$I  detailed Shell Model calculations 
are not available and we use the Odd Group Model \cite{ref:OGM} where only the 
unpaired proton contributes.
The resulting axial-vector current coefficients $S_{A}$ for pure bino
and pure photino (purity=1) obtained using the quark spin coefficients 
Global Fit-2 are listed in Tab.~\ref{tab:NaISn}.  
The SD cross section on $^{23}$Na is almost three times smaller 
than the values derived for $^{27}$Al with the OGM and the spin quark
coefficients Global Fit-2 (see Tab.~\ref{tab:Sa}). This is due to the
different total proton spins and nuclear spins.
In the case of $^{127}$I, the SD cross section evaluated with 
the Global Fit-2 is almost two orders of 
magnitude weaker than the SI part and can be neglected in the evaluation of 
$\sigma_{0}$.

The bands of points in the total cross section of Fig.~\ref{fig:crossNa}
are due to the axial coupling of neutralinos with a large photino
components, for which the SI cross section is many order of magnitude weaker 
than the SD part.

\subsection{$^{73}$Ge and $^{117}$Sn}
$\quad$
We will discuss now the case of neutron-odd nuclei starting with
$^{73}$Ge (7.8$\%$ natural abundance).
Although the Ge dark matter detectors presently used \cite{ref:dmGe} are made 
of natural germanium which is mainly $^{74}$Ge (36.5$\%$), 
$^{72}$Ge (27.4$\%$) and $^{70}$Ge (20.5$\%$), the
use of enriched $^{73}$Ge detectors is planned for the future 
\cite{ref:CDMS}.

In order to evaluate the SD cross section we use the 
quark spin coefficients $\Delta Q_{q}$ derived
in Ref.~\cite{ref:nucspin2} (Global Fit-2) and
the results of the detailed Shell Model calculations reported
in Ref.~\cite{ref:shellge} for the nuclear physics.
In Fig.~\ref{fig:binoSDge} the SD cross section is plotted versus 
the bino purity for data sets B and D without grand unification constraints.
In both cases the highest SD cross section is obtained for pure higgsinos or
for mixtures with a substantial higgsino component. In the case of set B this 
is different from what we calculated for Al (see Fig.~\ref{fig:BinoSD}) where
the maximum SD cross section was obtained for gaugino-like neutralinos.
This is due to the different values of the axial coupling coefficient
of pure higgsino in the case of proton or neutron odd nuclei as discussed
in Sec.~\ref{subsec:SD}.

Since the contribution of the light $u$ and $d$ quarks is
neglected in the computation of the SI cross section as 
discussed in Sec.~\ref{subsec:SI}, the dependence of the 
axial coupling on the neutralino composition does not differ from the one
discussed for the proton-odd $^{27}$Al nucleus. 
In Fig.~\ref{fig:SISDge} the SI cross section is plotted versus the SD part
for data sets B and D without grand unification constraints.
The SI cross section is generally bigger than the SD part because 
the quite large value of the atomic number enhances the 
scalar coupling. For a large selection of neutralino compositions, the
SD part can then be neglected in the evaluation of the total cross section.
 
In Fig.~\ref{fig:SDge} the SD and SI cross sections
are plotted versus the neutralino mass for the combined sets A--E
without the grand unification constraints.
The bands of points in the low mass region of the SD cross section 
correspond to the contribution of pure photinos of sets A--C
with set A giving the highest contribution.
The structure due to pure bino is less evident than in the case of 
Al. It must be noted that the photino population decreases with increasing
values of $\tan\beta$ as discussed in Sec.~\ref{subsec:comp}.
In Fig.~\ref{fig:gegut} we plot the SD and SI cross sections for the
combined sets A--E with grand unification constraints.
The dependence of $\sigma_{0}$ on the implementation of the grand
unification constraints is the same as discussed for Al in Sec.~\ref{subsec:Al}.

Another material planned for dark matter detectors is Sn
\cite{ref:dmSSG}. Natural tin consists of different isotopes among
which the only odd-isotope with a relevant natural 
abundance is $^{117}$Sn (7.7$\%$). 
The axial-vector current coefficients for pure bino and pure photino
interactions on the neutron-odd isotope $^{117}$Sn are listed in 
Tab.~\ref{tab:NaISn}. The axial
coupling of pure binos and pure photinos evaluated with the OGM 
\cite{ref:OGM} and the Global Fit-2 is roughly three times larger
on $^{117}$Sn than on $^{73}$Ge mainly because of the different nuclear
spin $J$.
The SI part can be derived from the SI cross section evaluated for $^{27}$Al 
with appropriated scaling due to the different nuclear mass.
 The cross section $\sigma_{0}$ (SI+SD) on
$^{117}$Sn is plotted in Fig.~\ref{fig:totalsn} for the combined sets A--E
without grand unification constraints.
Due to the large value of the atomic number, the SI
cross section is generally bigger than the SD part and, for a large selection
of neutralino compositions, the axial coupling can be neglected in the
evaluation of $\sigma_{0}$.

\section{Prospects for detection of low mass neutralinos}
\label{sec:CRESST}
$\quad$
The present experimental limits for WIMP dark matter are given by 
Ge~\cite{ref:dmGe} and Si~\cite{ref:dmSi} ionization detectors 
and NaI scintillators~\cite{ref:dmNaI}.
Due to their relatively high recoil-energy threshold, such detectors
are not sensitive to low mass ($M_\chi < 10$\,GeV) dark matter particles. 

In order to explore the low mass window, a dark matter search experiment 
(CRESST) based on the use of a sapphire (Al$_2$O$_3$) cryogenic detector 
with a low energy threshold is currently under preparation
\cite{ref:CRESST}.
The first stage of this experiment (CRESST-1) will use 
a 1\,kg sapphire detector with an expected energy threshold of 0.5\,keV 
and a full width half maximum energy resolution $\Delta E$=200\,eV.

We perform a rough estimate of the CRESST-1 sensitivity to neutralino
dark matter assuming a flat distribution of the radioactive background at
1 count/keV/kg/day.
A discussion of different radioactive background sources can be found in 
Ref.~\cite{ref:CRESST}. For each dark matter particle mass, 
 the simulated background data is fitted with the sum 
of a flat component for the background and the calculated recoil energy 
spectrum for $^{27}$Al. The exclusion limit is defined as the cross section
for which the fit gives a $\cal X$$^{2}$ value corresponding to
the 90$\%$ confidence level.
The expected recoil energy spectrum is calculated using an
exponential form factor to account for the loss of nuclear coherence at high
momentum transfer \cite{ref:OGM}. 
We use the same exponential form factor for both SD and SI interactions.
Due to the small nuclear radius the effect of the nuclear 
form factor on the shape of the recoil energy spectrum is
negligible for Al.
We use a Maxwell-Boltzman velocity distribution for the dark matter halo with
an average velocity of 270\,km/s and an upper cutoff at the escape velocity
of 575\,km/s in the rest frame of our galaxy.
The local halo density depends on the model
used to describe the structure of our galaxy with a resulting 
uncertainty in the detection rate of roughly a factor of two \cite{ref:Griest}.
In this work the density of dark matter particles is assumed to be 
0.3\,GeV/cm$^{3}$.

In Fig.~\ref{fig:crossCRESST} the cross section $\sigma_{0}$ (SI + SD),
calculated without the  grand unification constraints for the combined 
data sets A--E, is compared to
the rough estimate of the expected sensitivity of the first stage of the 
CRESST experiment assuming  a measurement time of 1 year. 
The expected low energy threshold of the cryogenic detector used in the
CRESST experiment combined with the low atomic number of Al provide an 
appreciable sensitivity to dark matter particles starting at $\sim$1\,GeV.
It is important to note that $^{16}$O does not contribute to the SD cross
section on sapphire because of the even number of nucleons. 
Due to the small nuclear mass, the SI cross section on $^{16}$O is roughly an 
order of
magnitude lower than on $^{27}$Al and can be neglected in the evaluation
of the sensitivity of a sapphire detector.

For completeness we have plotted in Fig.~\ref{fig:conver} the 
normalized interaction rate in units of counts/kg/day/pbarn that is, we show
the interaction rate assuming a cross section of 1\,pbarn.
The measured rate in a given detector will be less than the interaction 
rate, due to threshold and efficiency effects. 
The interaction rate is evaluated  using the analytical expression derived in
Ref.~\cite{ref:gabutti} with a zero energy threshold\footnote{We use 
the expression given for $(Rate_{c})$}.
The variation
with mass reflects the change in the dark matter flux combined with the 
effect of the nuclear form factor which is relevant for scattering of 
heavy neutralinos on large nuclei.
The variation with nucleus reflects the difference in the number of target
nuclei per unit weight.
In the case of $^{73}$Ge and $^{117}$Sn the interaction rate is calculated 
for a 100$\%$ enriched detector. 
We have separated the contribution of $^{23}$Na and $^{127}$I to the total
interaction rate of NaI detectors (50$\%$ Na and 50$\%$ I atoms).
In the case of sapphire detectors we have considered only the 
rate due to neutralino interactions on the $^{27}$Al content of sapphire 
since the neutralino cross 
section on  $^{16}$O is negligible.

A final question concerns how the counting rate for a given neutralino
will vary when we change the target material. This is of interest
in connection with two points. One is the verification of a presumed
dark matter signal and the other is discovering the nature of the neutralino.
Should there be an indication of a dark matter signal in a detector it 
will be of course of the utmost importance to verify that it is a true
signal and not the result of noise or backgrounds of some kind.
An aid in doing this will be variation of the target material, leading to
changes which are not necessarily those of the noise or background.
Secondly, 
varying the type of nucleus can help to determine the character of the 
particle, to see if its interactions are dominantly SI or SD, if
it is stronger on neutrons or protons and so forth, thus helping to pin
down the type of neutralino. Disentangling and understanding these various 
effects will be a question for much further detailed study. However, we can 
give a first impression on the basis of the calculations in this
paper by presenting examples with some of the various candidate
neutralinos that we have found in our study.

In Fig.~\ref{fig:fig24} we show the dependence of the 
cross section $\sigma_{0}$ for different 
neutralino compositions on various materials in the case of 
set B with $\tan\beta$=2, M$_{A}$=50\,GeV and the Higgs mixing angle 
$\alpha=-1.24$. We use Global Fit-2 for the quark spin coefficients and
the OGM for the total proton and neutron spins for Na, I and Sn. For
Al and Ge we use the shell model as discussed in Sec.~\ref{subsec:SD}.
The highest cross section is given by zino-higgsino mixtures
since pure higgsino and gaugino states do not have SI interactions, which
increase with $A^{2}$. For other examples, which we take to be pure 
higgsino or gaugino, we show no entries for $^{74}$Ge since it
has spin zero and thus there is no SD interaction. 
For pure higgsinos we have higher cross sections on neutron-odd nuclei,
 as discussed in Sec.~\ref{subsec:SD}. 
The cross section for pure higgsinos and pure
photinos on proton-odd nuclei tend to have roughly the same value.
In the case of pure higgsino or gaugino states, 
$\sigma_{0}$ does not scale with $A^{2}$ since
the SI part is absent and the dependence on the materials is due
to the nuclear model.
It should be stressed that Fig.~\ref{fig:fig24} represents
 only  a particular case.
As discussed in the previous sections, the cross section on nuclei
depends not only on the neutralino composition but also 
on the squark and Higgs masses and $\tan\beta$.

\section{Summary}
$\quad$
The minimal supersymmetric standard model as the 
low-energy effective theory 
has been used to explore the most general scenario for neutralino dark matter
with the present accelerator data and the requirement
0.025$< \Omega h^{2} <$1 as the only constraints without further theoretical 
assumptions. 
In order to obtain limits on the neutralino mass which are model-independent,
we scanned the parameter space randomly choosing the values of the free
parameters of the model using  different sets for the squark, 
slepton and CP-odd Higgs masses.

We have found an absolute lower bound for the mass of the cosmologically 
relevant neutralinos  of $M_{\chi} >$ 2\,GeV. The cosmologically relevant 
neutralinos are 
dominantly gauginos with an higgsino component coming in for masses 
above 30\,GeV.
We performed a similar analysis including the grand unification constraints
and obtaining dark matter neutralinos with a lower bound on the mass of 20\,GeV.
  
For the cosmologically relevant neutralinos, 
we have calculated the spin dependent (SD) and spin independent (SI)
cross sections on different nuclei under consideration for dark matter
detection. We have found a 5 orders of magnitude spread
in the values of the SD cross section
for the neutralino masses and composition considered in our analysis.
The spread is even more pronounced in the case of the SI cross section.

We have discussed the dependence of the SD cross section on the nuclear
model and on the nucleon quark spin content. We have used the
Odd Group Model and the detailed Shell Model calculations to evaluate
the total proton and neutron spins and  have found that
the uncertainty resulting from the choice of the nuclear model is within
a factor of two for the quark spin models with a non zero strange quark
contribution. The naive Quark Model, where only up and down quarks are
considered, gives the smallest cross section on neutron-odd nuclei
and the highest axial coupling for proton-odd nuclei.

For our selected neutralinos, the contribution of the photino component
to the SD cross section is
relevant only for small $\tan\beta$ because the photino population
is strongly reduced at large $\tan\beta$.
The highest SD cross section is obtained for the pure photinos and pure binos
of set A for which squark exchange gives a significant
contribution because of the small squark mass $M_{sq}$=45\,GeV.
For heavy squarks the highest SD cross section is
obtained for gaugino-higgsino mixtures with a substantial higgsino component
because squark exchange is suppressed and $Z$ exchange becomes dominant.
The higgsino contribution to the SD cross section is more relevant 
in neutron-odd nuclei.

In the case of $^{27}$Al, the SD and SI cross sections are comparable in
magnitude for a large selection of neutralino compositions. The SI cross 
section on $^{73}$Ge is generally bigger than the SD part because 
the larger value of the nuclear mass enhances the scalar coupling.
For a large selection of the parameter space, the SD part can be
neglected in the evaluation of the cross section $\sigma_{0}$.
This is true also for $^{127}$I where the SD cross section is almost two 
orders of magnitude smaller than the SI part and can be neglected.
Due to the different values of the total proton spins and nuclear spins,
 the SD cross section on $^{23}$Na is almost
three times smaller than on $^{27}$Al.

The SD cross section increases with the increasing higgsino component due to 
the $Z$ exchange contributions. SI cross sections are large for neutralinos
which are zino-higgsino mixtures. Large cross sections on nuclei correspond to
small relic abundance, close to our lower bound $\Omega h^2 > $ 0.025.

The neutralino cross section on $^{27}$Al was compared with a rough estimate
of the expected sensitivity of the first stage of the CRESST which 
uses a low energy threshold sapphire cryogenic detector.
We have shown that the expected sensitivity to low mass dark matter particles 
of the CRESST experiment can be useful in probing the light neutralino scenario 
($M_{\chi} >$ 2\,GeV) predicted by the present analysis based on the 
unconstrained minimal supersymmetric standard model.

\section*{Acknowledgement}
$\quad$
 We would like to thank
M.~Ted~Ressel from CalTech and A.~Bottino and S.~Scopel from the University of 
Turin for helpful discussions on some aspects of neutralino detection.
We would like to acknowledge helpful discussions with F.~Pr\"obst from the
Max-Planck-Institut f\"ur Physik.
One of us (M.~O.) would like to acknowledge the support of the
Polish Committee of Scientific Research.

\newpage
\section*{Figure captions}
\newcounter{fig}
\begin{list}{Fig. \arabic{fig}:}{\usecounter{fig}}
\item \label{fig:M2mu}
Scanned parameter space $a)$ without and $b)$ with grand unification
constraints. The existing limits from accelerator experiments are
incorporated. The contour line corresponds to the lower limit on the
chargino mass $M_{\chi^{\pm}} = 45$\,GeV evaluated for $\tan \beta$=1 (solid
lines) and $\tan \beta$=50 (dotted lines).
The horizontal line at $M_{2}$=50\,GeV in the GUT case corresponds to the
experimental bound from the gluino mass.
The upper left and right corners in the GUT case are empty because
in these regions $M_{\chi}\ge$100\,GeV.

\item \label{fig:omega}
Relic abundance versus neutralino mass in the parameter region allowed by the
experimental constraints for the four sets B--E listed in Tab.~\ref{tab:set}.
The plot for set A is very similar to that for B.
The cosmologically interesting relic abundance $0.025 < \Omega h^2 < 1$
is within the two horizontal lines.
In the case of sets A--C, the upper bound in the neutralino mass is
given by the slepton mass since we require $M_{\chi}\le M_{sl}$.

\item \label{fig:omegaGUT}
As in Fig.~\ref{fig:omega} with the grand unification constraints.

\item \label{fig:purity}
Gaugino purity $Z_{11}^{2}+Z_{12}^{2}$ $a)$ and bino purity $Z_{11}^2$
$b)$ for neutralinos in the cosmologically interesting region.
The data sets A--E are combined.
The points at bino purity $Z_{11}^2 = \cos^{2} \theta_{W} =
0.77$ correspond to pure photinos.

\item \label{fig:purityGUT}
As in Fig.~\ref{fig:purity} with the grand unification constraints.

\item \label{fig:Z3Z4}
Higgsino composition for the cosmologically interesting neutralinos
without grand unification constraints. The data sets A--E are combined.
The points with $Z_{13}^{2}+Z_{14}^{2} = 1$ correspond to pure higgsinos.

\item \label{fig:omegapur}
Relic abundance versus bino purity $Z_{11}^2$ for data sets B and D without
grand unification constraints.
The cosmologically interesting relic abundance $0.025 < \Omega h^2 < 1$
is within the two horizontal lines.

\item \label{fig:omegatan}
Bino purity $Z_{11}^2$ versus $\tan\beta$ for the cosmologically relevant
neutralinos of data sets B and D without grand unification constraints.

\item \label{fig:SISD}
SI versus SD cross sections 
on $^{27}$Al for the cosmologically relevant neutralinos
of data sets A, B, C and D without grand unification constraints.

\item \label{fig:BinoSI}
SI cross section on $^{27}$Al versus bino purity $Z_{11}^2$
for the cosmologically relevant neutralinos of data sets B and D with
$\tan\beta<$2 without grand unification constraints. The points near
$Z_{11}^2$=$\cos^{2}\theta \simeq$ 0.77 correspond to neutralinos with
a large photino component.

\item \label{fig:SItanbeta}
SI cross section on $^{27}$Al versus $\tan\beta$ 
for the cosmologically relevant neutralinos of data sets
B and D without grand unification constraints.

\item \label{fig:crossAl}
SD $a)$ and SI $b)$ cross sections on $^{27}$Al for the cosmologically
relevant neutralinos obtained without grand unification constraints.
The data sets A--E are combined.

\item \label{fig:photcross}
SD cross section on $^{27}$Al for the cosmologically
relevant photinos and binos with purity $>$0.99.
The contribution of the different sets A, B and C is shown.
For each set, the cross section of pure photino is $\sim$2 times larger than
for pure bino.

\item \label{fig:BinoSD}
SD cross section on $^{27}$Al versus bino purity
for the cosmologically relevant neutralinos of
 data sets B and D without grand unification constraints.

\item \label{fig:crossGUT}
SD $a)$ and SI $b)$ cross sections on $^{27}$Al for the
cosmologically relevant neutralinos obtained for the combined data sets A--E
with the grand unification constraints.

\item \label{fig:crossNa}
Cross section $\sigma_{0}$ (SI+SD) on $^{23}$Na and $^{127}$I
for the cosmologically relevant
neutralinos evaluated without grand unification constraints.
The sets A--E are combined.

\item \label{fig:binoSDge}
SD cross section of the cosmologically relevant neutralinos
on $^{73}$Ge versus bino purity for data sets B and D
without grand unification constraints.

\item \label{fig:SISDge}
SI cross section versus the SD part on $^{73}$Ge for the cosmologically
relevant neutralinos
of data sets B and D without grand unification constraints.

\item \label{fig:SDge}
SD $a)$ and SI $b)$ cross section on $^{73}$Ge versus the neutralino mass
for the cosmologically relevant neutralinos of combined data sets A--E
without grand unification constraints.

\item \label{fig:gegut}
As in Fig.~\ref{fig:SDge} with  grand unification constraints.

\item \label{fig:totalsn}
Cross section $\sigma_{0}$ (SI+SD) of the cosmologically relevant
neutralinos on $^{117}$Sn for the combined sets
A--E  without grand unification constraints.

\item \label{fig:crossCRESST}
Cross section $\sigma_{0}$ (SI+SD) on $^{27}$Al for the cosmologically
relevant neutralinos of data sets A--E without grand unification
constraints.
A rough estimate of the sensitivity of the first phase  of the CRESST
experiment is also shown for a measurement time of 1 kg-year assuming
a flat radioactive background of 1 count/kg/keV/day and a detector energy
threshold of 500 eV with a full width half maximum energy resolution
$\Delta E$=200\,eV.

\item \label{fig:conver}
Normalized interaction rate, in units of count/kg/day/pbarn, showing form
factor and dark matter flux effects. The rate is evaluated
with a zero energy threshold.
Solid line: rate in a sapphire detector due to interactions on $^{27}$Al.
Dashed lines: rate in a NaI detector due to
interactions on $^{23}$Na and on $^{127}$I respectively.
Dotted lines: rate in 100$\%$ enriched $^{73}$Ge (upper) and
$^{117}$Sn (lower) detectors.

\item \label{fig:fig24}
Cross section $\sigma_{0}$ (SI+SD)
normalized to the reduced mass squared for different
materials and neutralino compositions.
Solid triangles: zino-higgsino mixture with 40$\%$ zino composition.
Open triangles: pure higgsino. Solid circles: pure bino.
Open circles: pure photino. The cross sections are calculated for
set B with $\tan\beta$=2, $M_{A}$=50\,GeV and $\alpha=-1.24$.
As can be seen in Fig.~\ref{fig:purity},
zino-higgsino mixtures and pure
higgsinos are cosmologically relevant above $\sim$20\,GeV, while
binos and photinos are
cosmologically relevant above $\sim$2\,GeV (without grand unification
constraints).
\end{list}

\newpage
\section*{Tables}
\begin{table}[h]
\begin{center}
\caption[]{ \label{tab:range} \small Ranges used for the free parameters of 
the unconstrained MSSM. Calculations with the grand
unification constraints are performed using the same ranges but with
$M_1/M_2$ = 0.5 and $M_2 \ge$ 50\,GeV.\\}
\begin{tabular}{|c|ccc|}\hline
            & min. & max. &  \\ 
\hline\hline
$M_2$       & 1    & 1000 & GeV \\
$M_1/M_2$   & 0    & 1    & \\
$\mu$       &-1000 & 1000 & GeV \\
$\tan\beta$ & 1    & 50   & \\ \hline
\end{tabular}
\end{center}
\end{table}

\begin{table}[h]
\begin{center}
\caption[] { \label{tab:set} \small
Masses in GeV of the squark $M_{sq}$, slepton $M_{sl}$, and CP-odd Higgs $M_A$ 
used to compute the neutralino relic abundance and the cross section on 
nuclei for our different trial sets.\\} 
\begin{tabular}{|c|ccc|} \hline
set & $M_{sq}$ & $M_{sl}$ & $M_A$  \\ \hline \hline
A   &45&45&25-70\\
B   &100 &45&25-70\\
C   &150 &90&25-70\\
D   &200 &200&200\\ 
E   &1000 &1000&1000\\ \hline
\end{tabular}
\end{center}
\end{table}

\newpage
\begin{table}[h]
\begin{center}
\caption[]{ \label{tab:DQ} \small Axial coupling coefficients $a_{p(n)}$ for
pure photino and pure bino evaluated from Eq.~(\ref{eq:ap})
using different predictions for the quark spin coefficients $\Delta Q_{q}$.
The coupling coefficients are shown relative to $M_{W}$/$M_{sq}$=1.\\}
\begin{tabular}{|cc|cccc|}\hline
        &                & NQM   & EMC   & Global Fit-1 & Global Fit-2 \\ 
\hline\hline 
        & $\Delta Q_{u}$ & 0.93  & 0.78  & 0.80 & 0.82 \\
        & $\Delta Q_{d}$ & -0.33 & -0.5  & -0.16 & -0.44 \\
        & $\Delta Q_{s}$ & 0     & -0.16 & -0.13 & -0.11 \\ \hline\hline
photino & $a_{p}$     & -0.175  & -0.127 & -0.135 & -0.141 \\
        & $a_{n}$     & 0.020  & 0.071  & 0.060 & 0.054\\ \hline
bino    & $a_{p}$     & -0.119   & -0.084  &  -0.089 & -0.094\\
        & $a_{n}$     & 0.008   & 0.045  & 0.037 & 0.033\\ \hline 
\end{tabular}
\end{center}
\end{table}

\newpage
\begin{table}[h]
\begin{center}
\caption[]{ \label{tab:Sa} 
\small Axial-vector current coefficients $S_{A}$ 
 for pure photino and pure bino (purity=1) interactions on the
proton-odd nucleus $^{27}$Al and on the neutron-odd nucleus $^{73}$Ge.
The coefficients $S_{A}$ are evaluated using different 
nuclear models and the nucleon quark spin content coefficients listed in
Tab.~\ref{tab:DQ}. The $<S_{p(n)}>$ values used for the Shell Model are 
unquenched in the case of Al \cite{ref:MICA} and quenched 
(Q=0.833) in the case of Ge \cite{ref:shellge}. 
The coupling coefficients are evaluated relative to $M_{W}$/$M_{sq}$=1.\\}
\begin{tabular}{|c|cc||cc|}\hline
                             & $^{27}$Al (J=5/2) &  & $^{73}$Ge (J=9/2)& \\
\hline\hline
                             & OGM   & Shell Model & OGM   & Shell Model \\ 
\hline 
           $<S_{p}>$         & 0.25  & 0.3430 & 0 & 0.011 \\
           $<S_{n}>$         & 0     & 0.0296 & 0.23 & 0.468 \\ \hline\hline
$S_{A}$ for pure photino                    &       &  & & \\ \hline
NQM               & 0.0215 & 0.0396  &0.0002 & 2.9$\cdot$10$^{-8}$\\ 
EMC               & 0.0113 & 0.0193 & 0.0026 & 0.0058 \\ 
Global Fit-1      & 0.0127  & 0.0221 & 0.0018 & 0.0036 \\ 
Global Fit-2      & 0.0139  & 0.0245 & 0.0015 & 0.0026 \\ \hline \hline
$S_{A}$ for pure bino                        &       & & & \\ \hline
NQM            & 0.0099 & 0.0184 & 3.3$\cdot$10$^{-5}$ &
5.5$\cdot$10$^{-5}$\\
EMC            & 0.0049 & 0.0084 & 0.0011 & 0.0023 \\
Global Fit-1  & 0.0056 & 0.0098  &0.0007  & 0.0013 \\
Global Fit-2  & 0.0062 & 0.0110 & 0.0006 & 0.0009 \\ \hline
\end{tabular}
\end{center}
\end{table}

\newpage
\begin{table}[h]
\begin{center}
\caption[]{ \label{tab:NaISn} 
\small Axial-vector current coefficients $S_{A}$ 
 for pure photino and pure bino (purity=1) interactions on 
the proton-odd nuclei $^{23}$Na and $^{127}$I and on the neutron-odd nucleus 
$^{117}$Sn.
The coefficients are evaluated using the OGM for the nuclear structure
and the nucleon quark spin coefficients derived from a fit
of the present data on polarized lepton-nucleon scattering (Global Fit-2). 
The coupling coefficients are evaluated relative to $M_{W}$/$M_{sq}$=1.\\}
\begin{tabular}{|c|c||c||c|}\hline
                             & $^{23}$Na (J=3/2) & $^{127}$I (J=5/2)& 
$^{117}$Sn (J=1/2)\\
\hline\hline
           $<S_{p}>$         & 0.156  & 0.07 & 0 \\
           $<S_{n}>$         & 0     & 0 & 0.261  \\ \hline\hline
$S_{A}$ for pure photino   & 0.0065 &0.0011 &0.0048\\ \hline
$S_{A}$ for pure bino &   0.0029 &0.0005 &0.0018\\ \hline
\end{tabular}
\end{center}
\end{table}

\end{document}